\documentclass[11pt]{article}
\usepackage[titletoc,toc]{appendix}
\usepackage{url}
\usepackage{amssymb}
\usepackage[numbers,sort&compress]{natbib}
\usepackage{amsmath,amsfonts,graphicx,epsfig}
\usepackage{bm}
\usepackage{xcolor}

\def\bea{\begin{eqnarray}}
\def\eea{\end{eqnarray}}
\def\be{\begin{equation}}
\def\ee{\end{equation}}
\def\ba{\begin{array}}
\def\ea{\end{array}}

\begin{document}

\setlength\arraycolsep{2pt}

\renewcommand{\theequation}{\arabic{section}.\arabic{equation}}
\setcounter{page}{1}

\begin{titlepage}

\begin{center}

\vskip 1.0 cm

{\LARGE  \bf Initial conditions for the Galileon dark energy}

\vskip 1.0cm

{\Large
Cristiano Germani
}

\vskip 0.5cm

{Institut de Ci\'encies del Cosmos, Universitat de Barcelona\\ Mart\'i i Franqu\'es 1, 08028 Barcelona, Spain.
}

\vskip 1.5cm

\end{center}

\begin{abstract} 

Galileon models are among the most appealing candidates for Dark Energy. The reason is twofold: classically, they provide a tracking solution leading to an almost DeSitter space starting from very generic initial conditions in the deep radiation era. The second reason is the standard lore that Galileons are quantum mechanically stable. The latter property is certainly true in flat space-time, thanks to the non-renormalization theorems of galilean coupling constants. However, in a cosmological background, we show that quantum effects might dominate the classical trajectory. Assuming the radiation era to last at least up to the electroweak phase transition, the trajectory with initial conditions sitting on the tracker is ruled out. On the other hand, it is always possible to find a sub-space of initial conditions such that the dark energy solution approaches stably the tracker at late times. Fixing the value of initial conditions that best fit current data, and assuming that the galileon effective theory is valid up to the beginning of the radiation epoch, we found that the reheating temperature of the universe cannot be larger than $10^8 \ {\rm GeV}$. Reversing the argument, if dark energy will turn out to be in form of Galileons, the bounds by EUCLID on the initial conditions for these models will also be a bound on the reheating temperature of our Universe. 

\end{abstract}

\end{titlepage}

\setcounter{equation}{0}
\section{Introduction}

Classically, there is not any cosmological constant problem. Adding a cosmological constant to the Einstein-Hilbert action is just as good as adding matter fields. Because quantum field theory only treats the physics of energy transitions, its predictions are completely independent upon the absolute value of a possible vacuum energy. Gravity instead deals also with that and thus, quantum contributions to the cosmological constant (disregarded in quantum field theory) become relevant. Because gravity couple universally to matter fields, renormalised bubble diagrams from standard model particles generate a contribution to the cosmological constant that is several orders bigger than the value observed in current experiments (see \cite{martin} for a nice review of the problem). This is the cosmological constant problem.

Assuming that the bubble diagrams can be somehow cancelled out, still, observations require a form of energy that acts as an extremely small cosmological constant $\Lambda$. Of course, a possibility could be that the bubble diagrams are not quite canceled out. However, this miraculous un-perfect cancellation would require a huge amount of fine-tuning. 

Excluding then a cosmological constant, given the homogeneity and isotropy of our Universe, a simple alternative is that a scalar field provides the driving energy (dubbed Dark Energy (DE)) of the current Universe. 

It turned out that we have two main alternatives to do so. The first one is perhaps the less imaginative: The needed vacuum energy is given by a scalar field potential where, at least in the late time cosmology, the scalar moves very slowly (if moves at all). However, here we simply go back to the cosmological constant problem as $\Lambda$ is nothing else than the necessary uplift of the potential in order to have the correct vacuum energy.

The second alternative is perhaps more interesting: the vacuum energy is given by a kinetic instead of a potential energy. Because a canonical kinetic term alone cannot produce a very slowly varying vacuum energy, in order to mimic a cosmological constant we are bound to use theories with derivative couplings. Those theories are non-renormalizable in the sense of effective field theories (EFT). Therefore the question of whether cosmological background are quantum mechanically stable immediately arise.

If we want to avoid fine tunings, we then need that for any model of DE
\begin{itemize} 
\item a) a DeSitter space (where the Universe appears to go) is an attractor solution. In oder words, in order to be predictive, it is desirable that a large space of initial conditions for the DE scalar field leads to the same DeSitter solution. 
\item b) The classical solution that is tracked to the DeSitter vacuum must be (quantum mechanically) stable.
\end{itemize}

The Authors of \cite{cosmogal} have shown that for the specific model of a covariant Galileon field \cite{vikman} (shortly Galileon) the property a) is satisfied with the obvious condition that the mass scales involved in the specific model are parameterised by the cosmological constant today.

As already noticed, a renormalizable massless scalar field can only produce a solution that is diluting with the Universe expansion. Thus, in order to obtain an energy density that mimics a cosmological constant, one needs a trajectory dominated by the non-renormalizable interactions. In the last few years, it has been shown that this is not necessarily a quantum mechanical disaster. In fact, by considering the properly canonically normalised scalar field, it is possible to prove that background effects may ``unitarize" the system. What we mean here is that background effects rise the cut-off of the theory so that scattering amplitudes of fluctuations, with physically relevant wavelength, are within the unitary regime of the theory, therefore, in this case, they will not spoil the background solution. This was originally observed, in the framework of inflation, for Higgs inflation \cite{higgs} and variants. In particular in the so-called new Higgs inflation \cite{new}, the unitarization was morally the same as in Galilean models of DE during the DeSitter phase.

In this paper however, we attack the problem of quantum stability from a different perspective. We ask whether the background itself is quantum mechanically stable. In other words we will consider the infinitely long wavelength quantum corrections to the tree-level Galiean Lagrangian and check whether they will destabilise the classical solution. In fact, even assuming that quantum corrections in the DeSitter era are under control, one may ask the question if they are along the whole tracking solution. 

Galilean theories are interesting benchmark of all derivatively coupled theories because of their conserved currents and non-renormalization theorems. Indeed, they are the most quantum mechanically stable theories among the derivatively coupled ones. The reason is that in flat space, it has been proven that loop corrections to the Galileon action can only produce higher derivatives operators with respect to the ones appearing in the tree-level Lagrangian \cite{tetradis} (non-renormalization theorem \cite{nonr}), thus ``slow" trajectories are quantum mechanically stable. Although, as we checked, the non-renormalization theorem is also true in a Friedman-Robertson-Walker (FRW) metric, give the fact that higher derivatives there are covariant, Hubble scale factors might destabilise even ``slow" trajectories. 

 Let us over-simplify to give the feeling of how quantum corrections can become important. Consider a second order derivative interacting theory. Suppose that loop corrections generate only higher (covariant) derivative operators (as in the Galileon), for example $\sim \phi\frac{\square^2}{\mu^2}\phi$, where $\mu$ is a mass scale. In flat space, this operator would only produce higher partial derivatives of $\phi$. Then, as long as one moves in a slow trajectory, quantum corrections are negligible.

However, on a FRW background, one would also have terms like $\sim \phi\partial_t\left(\frac{H^2}{\mu^2}\right)\dot\phi$, where $H=\frac{\dot a}{a}$ and the FRW metric $ds^2=-dt^2+a^2 dx\cdot dx$ has been used. Close to a DeSitter solution, $\frac{H_{dS}^2}{\mu^2}$ is a slowly varying function and thus quantum corrections are again negligible on a slow trajectory. However, for example during radiation, the same operator would produce a term like $\sim\phi\frac{H_{rad}^2}{\mu^2}H_{rad}\dot\phi$. In this case, if the Hubble scale $H$ becomes of order $\mu$, the ``higher derivatives" quantum corrections will compete with the canonical $\phi\square\phi$, and the classical trajectory will not longer be trustable \footnote{We remark that the above example should only be taken as figurative since only interaction terms will be in general generated and, in reality, large numerical factors will help the stability (we will see that the system will be stable up until well deep in the radiation era). Nevertheless, the above example completely capture the root of the quantum instability in the dark energy trajectory.}.

Considering the one-loop effective action in the semiclassical limit (to be defined later on), it turns out that, at least during radiation, the tracking trajectories are parameterised by the ratio $\Delta=\frac{H\dot\phi}{M^3}$. Here, $M$ is the mass scale appearing in the dark energy Lagrangians, $H$ is the Hubble constant and finally $\dot\phi$ is the time derivative of the dark energy scalar.  There are obviously two possible regimes a) $\Delta> 1$ and b) $\Delta\leq 1$, although, it has been shown by \cite{cosmobound} that only the case $\Delta\gg 1$ is phenomenologically viable.
Unfortunately, in the latter case, the effective action is generically non-polynomial in powers of $\Delta$ (and in turns of $\phi$). Thus, in the regime in which $\Delta\gg 1$, the quantum corrections to the equation of motion are generically quite hard to extract. 
Nevertheless, we could show that the assumption that quantum corrections to the tracking solution are small, puts a constraint to the initial value of $\Delta$ during the radiation epoch. Interestingly enough, the best fit value for the dark energy initial condition saturates the bound on the reheating temperature in order to have leptogenesis.

Finally, we also discussed the case $\Delta\ll 1$ and found that it is more affected by quantum corrections than the previous case and thus ruled out under the assumption of leptogenesis (or at least under the assumption that radiation lasted at least until $z\sim 10^{11}$). This case corresponds to the case in which the dark energy trajectory is already on the tracker.

\section{The tracking solution}

Here we shall quickly review the main features of the tracking solution. Interested reader will find the details in \cite{cosmogal}. Given a scalar field $\phi$, the covariant Galilean Lagrangians are \cite{vikman}
\begin{eqnarray}
& & {\cal L}_1=M^3 \phi\,,\quad 
{\cal L}_2=(\nabla \phi)^2\,,\quad
{\cal L}_3=(\square \phi) (\nabla \phi)^2/M^3\,, \nonumber \\
& & {\cal L}_4=(\nabla \phi)^2 \left[2 (\square \phi)^2
-2 \phi_{;\mu \nu} \phi^{;\mu \nu}-R(\nabla \phi)^2/2 \right]/M^6,
\nonumber \\
& & {\cal L}_5=(\nabla \phi)^2 [ (\square \phi)^3
-3(\square \phi)\,\phi_{; \mu \nu} \phi^{;\mu \nu} \nonumber \\
& &~~~~~~~+2{\phi_{;\mu}}^{\nu} {\phi_{;\nu}}^{\rho}
{\phi_{;\rho}}^{\mu} 
-6 \phi_{;\mu} \phi^{;\mu \nu}\phi^{;\rho}G_{\nu \rho} ]
/M^9\,,
\label{lag}
\end{eqnarray}
where a semicolon represents a covariant derivative, $M$ is a constant 
having a dimension of mass, and $G_{\nu \rho}$ is the Einstein tensor.

The full theory, including dynamical gravity, reads
\begin{equation}
S=\int {\rm d}^4 x \sqrt{-g}\,\left[ \frac{M_{\rm pl}^2}{2}R+
\frac12 \sum_{i=1}^5 c_i {\cal L}_i \right]
+\int {\rm d}^4 x\, {\cal L}_{matter}\,,
\label{action}
\end{equation}
where $M_{\rm pl}$ is the reduced Planck mass,
$c_i$ are constants and ${\cal L}_{matter}$ the matter Lagrangian.

In the tracking solution we will set to zero the tadpole, i.e. $c_1=0$. Although this is re-generated by quantum corrections as a function like $\phi f(R,\square R,\ldots)$, we can always tune (renormalize) these terms to be sub-leading. Note that, although this can be done to stabilise the tadpole for homogeneous backgrounds, perturbations might still be sensitive to these quantum corrections. Here, we will however only focus on the stability of the background solution.

To make this model appealing we will avoid any hierarchy of scales. This is also what it is expected within the spirit of effective field theory. Translated in terms of coupling constants it implies that $c_i={\cal O}(1)$.

Considering the covariant Galileon as the only source for Dark Energy, the Hubble equation can then be standardly written as (assuming flatness)
\be
\Omega_{matter}+\Omega_{DE}=1\ ,
\ee
where $\Omega_i=\rho_i/(3M_{\rm p}^2 H^2)$ and $\rho_i$ is the energy density of any fluid. The DE dominance $\Omega_{DE}=1$, implies that
\be
c_2 x_{dS}^2=6+9\alpha-12\beta\ ,\ c_3x_{dS}^3=2+9\alpha-9\beta\ ,
\ee
where $\alpha=c_4 x_{dS}^4$, $\beta=c_5 x_{dS}^5$, $x_{dS}=\dot\phi_{dS}/(H_{dS} M_{\rm p})$ and finally $f_{dS}$ are the functions evaluated at the DeSitter point. The DeSitter point, the attractor of the full autonomous system (including matter), is found to be for $\dot\phi_{dS}=const$.

To match with current observations, we fix $H_{dS}\equiv \Lambda\approx 10^{-60}M_{\rm p}$ and $M^3=M_{\rm p}\Lambda^2$. From now on, we will fix units $M_{\rm p}=1$. Then we have that $x_{dS}\approx 10^{60}\dot\phi_{dS}$. 

The condition that all coefficients $c_i={\cal O}(1)$ implies that $\dot\phi_{dS}={\cal O}(1)10^{-60}$. Thus we define $\dot\phi_{dS}\equiv \gamma H_{dS}$ where $\gamma={\cal O}(1)$. From the Hubble equation solved at the DeSitter point, one can always find the explicit expression $\gamma=\gamma(\frac{H_{dS}}{M},c_i)$. However, as it is not important for the following discussion, we will not make that explicit.

The tracking solution works as follows. Define the two parameters
\be
r_1\equiv\frac{\dot\phi_{dS}H_{dS}}{\dot\phi_i H_i}= \gamma\Delta^{-1}\ , r_2\equiv\frac{\dot\phi_i^4}{\dot\phi_{dS}^4}r_1^{-1}\ ,
\ee
where $f_i$ correspond to the initial values. Then the classically stable tracking solutions are found for $0\leq r_2\leq 1$ and $r_1\lesssim 2$ thus $\dot\phi_i\lesssim \dot\phi_{dS}$ \cite{cosmogal}. In other words, for any initial conditions given after inflation (in the deep radiation era) within the allowed ranges of $r_1$ and $r_2$, the full system consisting in dark energy and matter fluids, will eventually evolve from the initial radiation to matter era until approaching the De Sitter stage at late times. 

Given the fact that $\dot\phi_{dS}H_{dS}=\gamma M^3$ we see that there is only a small parameter space such that $\Delta\leq 1$. 

\section{The one-loop effective action: generalities}
In this paper we will be interested in the quantum stability of the tracking solution. In this respect we will only consider the infinitely long wavelength limit of the effective one-loop action of the scalar field. This is pretty much the same philosophy used in the calculation of the effective potential for a canonical scalar field.

Neglecting quantum gravity corrections, we will be interested in whether the effective equation of motions for the scalar field are quantum mechanically stable. In other words, we will consider the usual semiclassical limit of gravity where, while the gravitational field is classical (quantum corrections are Planck suppressed), matter fields are quantum mechanically corrected. In addition, since we will be interested in the radiation epoch where the DE fluid is negligible with respect to the other fluids, we can consider the DE as a test field there.

Using a background field method, we will split the dark energy field into a time dependent vacuum expectation value $\phi(t)$ (we will only consider long wavelength) and quantum fluctuations $\delta\phi$
\be
\hat\phi(t,\vec{x})=\phi_0(t)+\delta\phi\ .
\ee
Note that this splitting should not be confused with the classical scalar fluctuations that involve a gauge issue on the gravity sector. This splitting simply means that we will integrate, in the path integral, over all possible paths that are perturbatively close to the classical solutions and are parameterised by the deviation from the classical solution ($\delta\phi$). Classical perturbations are instead {\it the} classical solution.

What we meant here as a classical solution is the saddle point of the effective action. However, as we want to test the hypothesis that the tracking solution of DE is quantum mechanically stable, we will assume that the classical solution will be dominated by the saddle point of the tree-level Lagrangian.

At linear order, the quadratic part of the quantum fluctuations calculated from the action \eqref{action} will generically be (we will give details later on)
\be
S_{\delta\phi}=\frac{1}{2}\int d^4 x \delta\phi\bigtriangleup \delta\phi\ ,
\ee
where $\bigtriangleup=\bigtriangleup(\dot\phi,\ddot\phi,H,a,M,c_i)$ is a differential operator. The one loop effective action is calculated as follows:
\be
Z=\exp(\frac{i}{\hbar}S)\int {\cal D}\delta\phi_c e^{\frac{i}{\hbar}S_{\delta\phi}}\ ,
\ee
where $S$ is the tree-level action and $\delta\phi_c$ is the canonicalised fluctuation.

The quantum contribution to the effective action is then
\be
W=\ln \frac{Z}{Z_0}\ ,
\ee
where $Z_0$ are background independent constants. Since $S_{\delta\phi}$ is quadratic (we are only considering one-loop corrections) that can be integrated ``exactly" obtaining
\be
W=\frac{1}{2}\ln\det\bigtriangleup=\frac{1}{2}\int d^4x \sqrt{g}\ln(\bigtriangleup)\ .
\ee
In the stable case in which no gradient and ghost instability are present, given the gravitational background symmetries, the operator $\bigtriangleup$ acting on the canonical fluctuations will be a hyperbolic operator of the form
\be
\bigtriangleup\delta\phi_c=\delta\ddot\phi_c+B(t)\delta\dot\phi_c-C(t)\sum_{i}\partial_i\partial_i\delta\phi_c+m^2(t)\delta\phi_c\ ,
\ee
where the time-dependent coefficients will be worked out later on. In particular, in absence of gradient instability, $C(t)>0$. 

To calculate the determinant of $\bigtriangleup$ we will use the heat kernel technique. Assuming no gradient instabilities, in order to do so, we will firstly need to work with an elliptic differential operator. Background symmetries help us to transform the hyperbolic operator into an elliptic one, simply Wick rotating the space-coordinates $\partial_i\rightarrow i\partial_i$. Thus, we consider the ``Wick" rotated operator 
\be
\bigtriangleup_W\delta\phi=\delta\ddot\phi_c+B(t)\delta\dot\phi_c+C(t)\sum_{i}\partial_i\partial_i\delta\phi_c+m^2(t)\delta\phi_c\ .
\ee
Secondly, because techniques to calculate the heat kernel in curved space do exist, we will need to re-interpret $\bigtriangleup_W$ as a covariant operator in some auxiliary space. However, that will not be enough. The operator $\bigtriangleup_W$ contains also an off-diagonal term ($B(t)$). In order to include this in the formalism, we will need to extend our theory including a gauge field associated to $B(t)$ \cite{vassilevich}. In this way, $\bigtriangleup_W$ can be re-interpreted as the following covariant (and gauge-invariant) operator \cite{vassilevich}
\be
\bigtriangleup_W\delta\phi_c=(g^{\mu\nu}D_{\mu}D_{\nu}-E)\delta\phi_c\ ,
\ee  
where $D_{\mu}=\nabla_\mu+A_\mu$. Here $\nabla_\mu$ is the covariant derivative associated to an auxiliary metric $g_{\mu\nu}$, $A_\mu$ is a gauge field and $E$ is an endomorphism associated to $A_\mu$. Since $\bigtriangleup_W$ only acts onto scalars, we also define the associated field strength to $A_\mu$ to be
\be
F_{\mu\nu}=2\partial_{[\mu}A_{\nu]}\ .
\ee
Having done these definitions, we already expect that the effective one-loop action will be a function of only covariant quantities in the auxiliary manyfold with metric $g_{\mu\nu}$, in other words, we already expect that the effective action will only be a function of curvatures $R_{\mu\nu\alpha\beta}(g)$, $E$, the field strength $F_{\mu\nu}$ and covariant derivatives of those \cite{zanusso}.

Let us define the heat kernel to be
\be
H(s)=e^{-s \bigtriangleup_W}\ ,
\ee
then we have that
\be\label{W}
W=-\frac{1}{2}\int^\infty_{1/\Lambda_c^2}\frac{ds}{s}\int d^4x \sqrt{g}H(s)\ ,
\ee
where a cut-off $\Lambda_c$ has been introduced to regularise the divergences.

The way the heat kernel is calculated is via the off-diagonal heat kernel defined as a solution of the heat equation
\be\label{heat}
(\partial_s+\bigtriangleup_x)H(t,x,y)=0\ ,
\ee
with boundary conditions 
\be
H(0,x,y)=\delta(x,y)\ .
\ee
Then the heat-kernel is $H(s)=H(s,x,x)$. We are now left to solve the differential equation \eqref{heat}. 

In flat space $\bigtriangleup=\delta^{\mu\nu}\partial_\mu\partial_\nu$ the heat kernel solution is a Gaussian
\be
H(x,y,s)\Big|_{flat}=\frac{\bar H}{(4\pi s)^2}e^{-\frac{(x-y)^2}{4s}}\ ,
\ee
where $\bar H$ is a constant amplitude fixed to be $1$ by boundary conditions. Since any manyfold is locally flat we can already guess that the solution will be similar to the flat one with the substitutions $(x-y)^2\rightarrow \sigma(x,y)$ \cite{zanusso}. Where, $\sigma(x,y)$ is half of the squared geodesic distance between $x$ and $y$ \cite{8-zan} in the auxiliary metric $g_{\mu\nu}$ and $\bar H\rightarrow \bar H(x,y,s)$ where $\bar H(x,y,0)=1$. With this, the usual way to solve the heat equation is in the adiabatic expansion, i.e. in series of $s$, assuming a convergence. We will not go into details here, they can be found for example in \cite{vassilevich,zanusso,barvinski}. 

The integral in $s$ in \eqref{W} is also divergent at the infra-red. Of course, this divergence cannot be physical. If $\delta\phi$ had a mass 
$m$ then the heat kernel including also the mass, would be
\be
H(x,y,s)=\frac{\bar H(x,y,s)}{(4\pi s)^2}e^{\frac{-\sigma(x,y)}{4s}-s m^2}\ ,
\ee
then the one loop effective action could be trusted up to energy scales way below $m$ and the integral would converge. However, in our case $\delta\phi$ is of order of the Hubble scale $H$ and may even be tachyonic (for example in DeSitter background)\footnote{Note that this does not mean that the system is unstable. The tachyonic instability is usual in a DeSitter background and simply means that although the canonical scalar exponentially grow, this growing is tight to the exponential grow of the scale factor and the non-canonical scalar is simply frozen, i.e. the energy density is constant.}. Since we are precisely interested in that scales we will need to re-sum the logarithmic divergences. This, for $F_{\mu\nu}=0$, has been done in \cite{barvinski}. Has we shall soon prove, it is precisely this the case at our hands.

The effective action will contain terms that are subject to renormalisation conditions and some others that do not. We will here focus only on the terms that are independent on this arbitrariness. Note that, given the different functional form, the logarithmic corrections cannot cancel the finite terms. 

Then, at quadratic order, the quantum correction to the effective action (back to Lorentzian signature $d^3x\rightarrow -i d^3x$) will be
\be
W=\frac{1}{18}\frac{1}{16\pi^2}\int d^4x \sqrt{|g|}\left(ER-\frac{1}{6}R^2\right)\ ,
\ee
where $R$ is the Ricci scalar on the auxiliary metric $g_{\mu\nu}$.

\section{Quantum analysis}

The linear fluctuations of the dark energy are described by the equation of motion (after Wick rotation of the spatial coordinates)
\be
D_2(t)\delta\ddot\phi+D_4(t)\delta\dot\phi+\frac{D_9(t)}{a(t)^2}\partial_i^2\delta\phi=0\ ,
\ee
where $D_i$ are background dependent coefficients and can be found in \cite{coeff} (although for convenience we will list them again in the appendix). We remind the reader that we are calculating {\it quantum} fluctuations of the scalar and neglect quantum fluctuations of the metric, i.e. we are in the standard semiclassical regime. Classical fluctuations instead couple to the metric fluctuations and cannot be integrated out.

First of all we need to canonically normalise the fluctuations. This generically improve quantum stability for the perturbations (see e.g. \cite{nico}). 

The above equation of motion comes from an action of the form
\be
S_{\delta\phi}=\frac{1}{2}\int d^4x a^3 D_2\delta\dot\phi^2+\ldots\ ,
\ee
the canonical fluctuations are then defined as $\delta\phi_c=a^{3/2}\sqrt{D_2}\delta\phi$ where, requiring absence of ghost instability, $D_2>0$. For simplicity of notation, and to make connection for example with the Mukhanov-Sasaki variable \cite{muk}, we will define the non-standard part of the normalisation $D_2\equiv N^2$. Because the action is linear $D_4=\frac{1}{a^3}\partial_t(a^3 N^2)$. By defining the effective mass
\be
m^2=-\left(\frac{9}{4}H^2+\frac{3}{2}\dot H+3H\frac{\dot N}{N}+\frac{\ddot N}{N}\right)\ ,
\ee
we have that the canonical fluctuation evolves as
\be\label{can}
\delta\ddot\phi_c+c_s^2(t)\partial_i^2\delta\phi_c+m^2(t)\delta\phi_c=0\ ,
\ee
where the ``sound speed" $c_s^2=\frac{D_9}{a^2 D_2}$.

\subsection{The auxiliary metric}

We need now to interpret the equation of motion \eqref{can} in terms of the auxiliary metric $g_{\mu\nu}$ and the gauge vector $A^\mu$.

It is immediately clear to see that the spatial part of the metric is diagonal and thus $g^{ii}=c_s^2$. At the same time, by construction (we are working with a canonical field) $g^{tt}=1$. With this in mind, $\sqrt{g}=c_s^{-3}$.

Considering that
\be
\square_g\delta\phi_c=c_s^3\partial_t(c_s^{-3})\delta\dot\phi_c+\ldots\ ,
\ee
and that the above term does not appear in the equation of motion \eqref{can} we will need to absorb it inside a gauge field. Defining the operator $D_\mu=\nabla_\mu+A_\mu$ we have that
\be\label{endo}
g^{\mu\nu}D_{\mu}D_\nu\delta\phi_c=\square_g \delta\phi_c+\nabla_\alpha A^\alpha\delta\phi_c+2A_\mu\nabla^\mu\delta\phi_c+A^2\delta\phi_c\ ,
\ee
with this we identify
\be
A^t=-\frac{1}{2}c_s^3\partial_t(c_s^{-3})\ ,
\ee
and $A_i=0$. Thus the gauge field (as we would expect) is only along the gauge direction and therefore $F_{\mu\nu}=0$. Now the ``mass" terms in \eqref{endo} and the physical mass $m$ have to be absorbed by the endomorphism $E$. We then finally get
\be
E=\nabla_\mu A^\mu+A^2-m^2\ .
\ee

\section{The $\Delta\leq 1$ case}

We have now all the ingredients to find the quantum part of the effective action $W$. This action will be populated by higher derivatives of the field $\phi$ as well as lower (up to second), as described in the introduction. The full action $W$ contains a large number of terms and it is in general not polynomial on $\phi$ (and its derivatives). 

In the  $\Delta\leq 1$ the tracking solution is basically already on the tracker during the radiation epoch. This case has been shown to fit badly the data in \cite{cosmobound} (see also \cite{pascoli}). Nevertheless, it is still an interesting case that might be taken as a benchmark for all derivatively coupled theories of DE.

Since here $\Delta\leq 1$, we can expand the action $W$ in powers of it (and derivatives). Although $\Delta$ can only be slightly smaller (or equal) than $1$ \cite{cosmogal}, and therefore a first order expansion might be really crude, our analysis at first order will be enough to make our point of quantum instability. 

As we discussed in the introduction, the largest contribution from $W$ comes from powers of $\dot\phi$ accompanied by powers of $H$ (i.e. by $\Delta$). The growth of the Hubble constant is then the key reason by which $W$ might eventually dominate over the tree-level action. 
Indeed, as we shall show, the coefficients multiplying the powers of derivatives of $\phi$ in the equation of motion of $W$, become larger than the similar ones obtained from the tree-level Lagrangian in the deep radiation era, at about $z_c\sim 10^{11}$. Thus, the classical theory cannot be trusted at least at that red-shift.

The ratio between the $\dot\phi^2$ coefficient in the equation of motion from $W$ ($C_2^W$) and the equivalent one coming from the tree-level Lagrangian ($C_2^{T}$) is found to be (in $M_{\rm p}=1$ units)
\be\label{calR2}
{\cal R}_2\equiv\frac{C_2^T}{C_2^W}=f(\alpha,\beta)\frac{H_{dS}^4}{H_{rad}^6}\ ,
\ee
where 
\be
f(\alpha,\beta)\equiv -\frac{35 (2+9\alpha-9\beta)^2}{24\pi^2(2+3\alpha-4\beta)^3}\ .
\ee
In general we also found that ${\cal R}_i\equiv\frac{C_i^T}{C_i^W}\propto\frac{H_{dS}^4}{H_{rad}^6}$, where $i$ is the power of $\dot\phi$ in the equation of motion.

In \eqref{calR2}, the loss of predictability, i.e. the dominance of quantum corrections, is for ${\cal R}_2\leq 1$. Using $\Lambda$CDM as a reference, by using Hi-Class \cite{hi}, one finds that \footnote{I would like to thank Emilio Bellini for this.} at the matter-radiation equality $z_{r}\sim \frac{1}{3}10^4$, the Hubble constant $H(z_{r})\sim 1.5\times 10^5 H_{dS}$. Therefore, in the deep radiation era, the Hubble constant will evolve like
\be
\frac{H_{rad}}{ H_{dS}}\sim 10^{-2} z^2 \ ,
\ee
so that (we remind that $H_{dS}\approx 10^{-60}$)
\be\label{calR}
{\cal R}_{2}\sim 10^{132} z^{-12}f(\alpha,\beta)\ .
\ee
The allowed values of $\alpha$ and $\beta$ for the tracking solutions are listed in \cite{cosmogal}. We have checked that, within this range, $f(\alpha,\beta)\leq {\cal O}(100)$, thus, in the best case, ${\cal R}_2\sim 10^{134} z^{-12}f(\alpha,\beta)$ .
The appearance of a factor $10^{132}$ seems to make impossible the domination of the loop corrections. However, since the red-shift appears as a large power too ($z^{12}$), for large values of $z$, but still during the radiation era, the red-shift overcomes the huge numerical factor. We then see that $10^{134}z^{-12}$ becomes of order one for a redshift or order $z_c\sim 10^{11}$. Thus, at least for redshift higher than $z_c$, i.e. way after a would be electroweak phase transition in the early Universe ($z_{ew}\sim 10^{15}$), the classical tracking solution is not reliable. 

\subsection{The $\Delta\gg 1$ case}

In this case, the most phenomenologically relevant, the largest power on field derivatives dominate during radiation epoch \cite{cosmogal}. To simplify the calculation, and make a clear point, we can then well approximate the theory during the radiation epoch by only considering the Lagrangian with higher powers of the field derivatives. Without loss of generality we can assume $\beta\neq 0$ \footnote{If $\beta=0$ the same conclusions will be qualitatively valid but parameterised by $\alpha$.}. In this case, the system is completely dominated by ${\cal L}_5$, therefore, the tree-level solution approximately satisfies the differential equation (see appendix)
\be\label{eqm0}
\ddot\phi=\frac{3}{4}H_{rad}\dot\phi\ .
\ee

Assuming that the tracking solution is the dominant solution of the system, by varying the effective one-loop action constructed only out of ${\cal L}_5$ and using \eqref{eqm0}, we found the correction to the tree-level equation of motion to be
\be
\Delta{\rm Eqm}\simeq 481\frac{H_{rad}^5}{\dot\phi}\ .
\ee
Note that because of the canonical normalisation of the fluctuations, the above result does not depend on $\beta$.

The variation of the tree-level action gives us instead the following equation of motion
\be
{\rm Eqm}={\rm Eqm}_1+\ldots
\ee
where 
\be
{\rm Eqm}_1=\frac{45\beta H_{rad}^4}{H_{dS}^6\gamma^5}\dot\phi^4
\ee
and the dots are the second order derivative terms. We can now compare, for example, the term ${\rm Eqm}_1$ with $\Delta{\rm Eqm}$: assuming that the tree-level Lagrangian dominates during radiation, classical stability implies ${\cal R}_b\equiv \frac{{\rm Eqm}_1}{\Delta {\rm Eqm}}\gg 1$. Plugging the numbers and considering that $\beta\sim {\cal O}(1)$, we have
\be
{\cal R}_b\sim 10^{130}r_1^{-5} z^{-12}\ ,
\ee
which is similar to the previous result but now with the help of a $r_1^5\ll 1$ factor in favour of stability. 

If we consider the deep radiation era, for example during the electroweak phase transition ($z\sim 10^{15}$ and temperature $T_{ew}\sim 100\ {\rm GeV}$), quantum correction are under control for $r_1\ll 10^{-10}$. 

The best fit for the initial value of $r_1$ with data has been found in \cite{cosmobound}. The analysis of \cite{cosmobound} is based on WMAP7 data and of course can be improved with the use of recent data (see e.g. \cite{pascoli}). However, as we are only interested in order of magnitudes, we will be content with the results of \cite{cosmobound}. Then, the order of magnitude best fit is for $r_1\sim 10^{-10}$ at $z\sim 4\times 10^8$. During radiation, it is easy to find that $r_1\propto a^{5/4}$ \cite{cosmobound}. For a large redshift $z$, the temperature of the Universe scales as $T\propto z$. Therefore we have, during radiation, 
\be
r_1(T)\sim 10^{-16}\left(\frac{\rm GeV}{T}\right)^{5/4}\ ,
\ee
leading to
\be
{\cal R}_b(T)\sim 10^{54}\left(\frac{\rm GeV}{T}\right)^{23/4}\ .
\ee
We see then that quantum corrections become of order one at $T\sim 10^9\ {\rm GeV}$, thus we expect a maximal reheating temperature of order $T_{rh}\sim 10^8\ {\rm GeV}$. 

One of the most natural candidates for baryogenesis is leptogenesis. In this scenario, the reheating temperature of our Universe must be larger or equal than $T_{rh}\sim 10^8\ {\rm GeV}$ \cite{lepto}. Thus, the best fit for galilean dark energy initial conditions, together with the requirement of a successful leptogenesis and a quantum stability of the tracking trajectory, point out a precise reheating temperature of order $T_{rh}\sim 10^8\ {\rm GeV}$. Whether dark energy is related or not to the physics of reheating and leptogenesis is however at this level only a numerical curiosity.

\section{Conclusions}

In this paper, we have discussed the quantum stability of the background solution for a generic Galilean Dark Energy model. In the semiclassical limit, we have calculated the term in the one loop action that is independent upon renormalization conditions and showed that
\begin{itemize}
\item In the case in which the DE scalar is on a tracking solution leading to the DeSitter tracker, we found the bounds that the initial conditions for the dark energy scalar must satisfy in order to follow a quantum mechanically stable trajectory. In particular, for the initial conditions favoured by comparisons with observations, the one loop effective action become dominant at a temperature of order $T\sim 10^9\ {\rm GeV}$. Interestingly, this point out that, if the dark energy field is fundamental and follows a classical tracking solution from after reheating to our epoch, and baryogenesis happened because of leptogenesis, the reheating temperature must be of order $T_{rh}\sim 10^8\ {\rm GeV}$.

\item In the case in which the DE scalar is instead on (or very close to) the tracker solution during radiation era (the case disfavoured by data) we showed that quantum corrections become dominant at redshift $z\sim 10^{11}$. Thus, at least at $z\sim 10^{11}$, the Galilean tracking solution is not trustable. Requiring at least an electroweak phase in our Universe implies that this region of initial conditions is ruled out.

\end{itemize}
The above results should be seriously considered as guide-lines to select the correct initial conditions for the numerical simulations of (Galielan) dark energy models. Those simulation will be soon probed by the European EUCLID experiment and therefore, if dark energy will turn out to be in form of Galileons, our analysis will provide an independent indirect measure of the reheating temperature of our Universe.

Finally, we would like to remark that the Galilean theories are benchmarks for any dark energy models dominated by derivative couplings. The reason is that Galileons are the most quantum mechanically stable among all those models, thanks to their non-renormalization theorems. 
Therefore, we expect that any generic scalar-tensor theory of dark energy (e.g. Horndeski theories \cite{hor}) should suffer from larger quantum instabilities than the ones discussed here. However, this last check has to be done case by case and it is therefore way beyond the scope of this work.

\section*{Acknowledgments} 

I would like to thank Emilio Bellini for the many discussions on the topic and A. O. Barvinsky for correspondence. I am supported by the Ramon y Cajal program and partially supported by the Unidad de Excelencia María de Maeztu Grant No. MDM-2014-0369 and FPA2013-46570-C2-2-P grant.

 \appendix
 \section{Background equations and scalar fluctuations}

The background equations for the Galileon field are
\begin{eqnarray}
0 &=& c_2\left[\ddot{\phi} + \dot{\phi}\theta \right] + \frac{c_3}{M^3} \left[ 4\ddot{\phi}\dot{\phi}\theta + 2\dot{\phi}^2\theta^2 + 2\dot{\phi}^2\dot{\theta}\right] \nonumber \\
&&  + \frac{c_4}{M^6}\left[  6\ddot{\phi}\dot{\phi}^2\theta^2 + 4\dot{\phi}^3\dot{\theta}\theta + 2\dot{\phi}^3\theta^3 \right] \nonumber \\
&& + \frac{c_5}{M^9}\left[ \frac{5}{9}\dot{\phi}^4\theta^4   +\frac{20}{9}\ddot{\phi}\dot{\phi}^3\theta^3  +\frac{5}{3}\dot{\phi}^4\dot{\theta}\theta^2 \right],
\end{eqnarray}
where the expansion rate $\theta=3H$.

The coefficients of the scalar fluctuations can be found in \cite{coeff} by setting all gravity perturbations to zero. We then have
\begin{eqnarray}
& &D_2=c_2-6c_3 H\dot{\phi}/M^3+54c_4 H^2 \dot{\phi}^2/M^6
-60c_5 H^3 \dot{\phi}^3/M^9\,,\\
& &D_4=3c_2H-6c_3 \dot{H} \dot{\phi}/M^3-18c_3 H^2 \dot{\phi}/M^3
-6c_3 H\ddot{\phi}/M^3+162c_4 H^3 \dot{\phi}^2/M^6+108c_4 H \dot{H}
\dot{\phi}^2/M^6 \nonumber \\
& &~~~~~~~~+108c_4 H^2 \dot{\phi} \ddot{\phi}/M^6-180c_5
(H^3 \dot{\phi}^2 \ddot{\phi}/M^9+H^2 \dot{H} \dot{\phi}^3/M^9
+H^4 \dot{\phi}^3/M^9)\,,\\ 
& &D_9=c_2-4c_3 H \dot{\phi}/M^3
-2c_3 \ddot{\phi}/M^3+26c_4 H^2 \dot{\phi}^2/M^6
+12c_4 \dot{H} \dot{\phi}^2/M^6
+24c_4 H \dot{\phi} \ddot{\phi}/M^6 \nonumber \\
& &~~~~~~~~-36c_5 H^2 \dot{\phi}^2 \ddot{\phi}/M^9
-24c_5 (H^3 \dot{\phi}^3/M^9
+H \dot{H} \dot{\phi}^3/M^9)\,.
\end{eqnarray}

\end{document}